\documentclass[12pt,a4paper]{iopart}

\usepackage{array}
\usepackage{graphicx}
\usepackage{dcolumn}
\usepackage{bm}
\usepackage{upgreek}
\usepackage{amssymb}
\usepackage{url}
\usepackage{xcolor}
\usepackage{cite}

\begin{document}

\title{Relation Between Photoionisation Cross Sections and Attosecond Time Delays}
\author{Jia-Bao Ji$^{1,*}$, 
        Anatoli S. Kheifets$^{2,\dag}$,
        Meng Han$^{3,\ddag}$, \\
        Kiyoshi Ueda$^{1,4,5,\P}$
        and
        Hans Jakob W{\"o}rner$^{1,\S}$}
\address{$^1$ Laboratorium f{\"u}r Physikalische Chemie, ETH Z{\"u}rich, 8093 Zürich, Switzerland}
\address{$^2$ Research School of Physics, The Australian National University, Canberra ACT 2601, Australia}
\address{$^3$ J. R. Macdonald Laboratory, Department of Physics, Kansas State University, Manhattan, KS 66506, USA}
\address{$^4$ Department of Chemistry, Tohoku University, Sendai, 980-8578, Japan}
\address{$^5$ School Physical Science and Technology, ShanghaiTech University, Shanghai 201210, China}

\ead{
\mailto{$^*$ jiabao.ji@phys.chem.ethz.ch},
\mailto{$^\dag$ a.kheifets@anu.edu.au}, \\
\mailto{$^\ddag$ mengh@phys.ksu.edu},
\mailto{$^\P$ kiyoshi.ueda@tohoku.ac.jp}, \\
\mailto{$^\S$ hwoerner@ethz.ch}
}

\begin{abstract}
Determination and interpretation of Wigner-like photoionisation
delays is one of the most active fields of attosecond
science. Previous results have suggested that large photoionisation
delays are associated with structured continua, but a quantitative
relation between photoionisation cross sections and time delays has
been missing. Here, we derive a Kramers-Kronig-like relation between
these quantities and demonstrate its validity for (anti)resonances. 
This new concept defines a topological analysis, which rationalises 
the sign of photoionisation
delays and thereby sheds new light on a long-standing controversy
regarding the sign of the photoionisation delay near the Ar $3s$ Cooper
minimum. Our work bridges traditional photoionisation spectroscopy
with attosecond chronoscopy and offers new methods for analysing and
interpreting photoionisation delays.
\end{abstract}

\maketitle

\section{Introduction}
Photoionisation is one of the fundamental processes that has been employed to reveal the
electronic structure and dynamics of matter. Traditionally,
photoionisation spectroscopy has been realised in the frequency domain
by measuring the yield and angular distributions of photoelectrons
\cite{becker1996vuv,Schmidt1997}. Newly developed laser-assisted
interferometric techniques expanded these studies into the time domain
and heralded the advent of attosecond science
\cite{corkum07a,RevModPhys.81.163}.  The methods of attosecond
streaking \cite{itatani02a,kienberger03a} and reconstruction of
attosecond beating by interference of two-photon transitions (RABBIT)
\cite{veniard96a,paul01a} were particularly instrumental in this new
field. With these metrologies, photoionisation dynamics becomes directly accessible on the
attosecond scale \cite{schultze10a,kluender11a}, yielding the
Wigner-like time delay
\cite{eisenbud1948formal,wigner1955lower,smith1960lifetime}, which
describes the phase variation over the photoelectron energy. When reviewing
the numerous measurements and calculations of photoionisation delays
that have been reported to date, it becomes apparent that large
photoionisation delays (in magnitude) are usually associated with
structures in the photoionisation continuum, such as Fano resonances
\cite{kotur2016spectral,gruson16a,zhong2020attosecond}, shape
resonances
\cite{huppert16a,baykusheva17a,heck2021attosecond,nandi20a,heck2022two, hammerland24a} and Cooper
minima \cite{schoun14a,alexandridi2021attosecond}. For example, in the case of CF$_4$,
photoionisation delays of up to $\sim$600~as have been measured and
shown to originate from trapping of the outgoing photoelectron in a
molecular shape resonance, whereby the trapping time is exacerbated by
a molecular cage effect \cite{heck2021attosecond}. Naturally, this
shape resonance also manifests itself as a local enhancement of the
photoionisation cross section, but the width of the associated
resonance cannot be trivially related to the trapping time, as
discussed in \cite{heck2021attosecond}. In the simple case of a single
continuum channel and single partial wave, the cross section of the shape resonance can be
related directly with the corresponding time delay
\cite{kheifets2023shape}. However, the interference between
resonant and non-resonant photoionisation channels causes a more
complex relationship between the time- and frequency-domain
manifestations of photoionisation dynamics
\cite{holzmeier2021influence}.  Do these results imply that no simple
relationship between these two fascets of photoionisation exists at
all?

In this work, we derive a surprisingly simple and general relationship
between the energy dependence of photoionisation cross sections and
attosecond photoionisation delays. We demonstrate the validity of this
relation for energetically confined (anti)resonances. In addition
to linking in a quantitative manner time- and frequency-resolved
photoionisation spectroscopies without restrictive assumptions, our
results introduce a topological analysis, which explains why and under
which circumstances the sign of photoionisation delays at
(anti)resonances can change from one calculation to another. Such
results have led to a notable controversy regarding the Cooper minimum (CM)
in the $3s$ continuum of argon, that is induced from the $3p$
continuum through inter-shell correlation. Whereas some calculations
have reported positive photoionisation delays near the $3s$ CM \cite{guenot2012photoemission,kheifets2013time,PhysRevA.91.063415}, other
calculations have reported negative delays
\cite{dixit2013time,pi2018attosecond}, whereas the authors of
  \cite{dixit2013time} later corrected their result and
  reported a positive delay \cite{PhysRevA.91.063415}. Our present
method shows that the local sign of these delays is related to the
topology of the complex-valued photoionisation matrix element,
i.e. its winding number in the complex energy plane. This explains
why, contrary to current belief, a quantitative reproduction of the
cross section is not sufficient to guarantee the accuracy of
calculated photoionisation delays. The topological analysis of the trajectories of the transition amplitudes is likely to facilitate the interpretation of
photoionisation delays in structured continua and thereby to become a
powerful and widely applicable approach for extracting the physics
responsible for non-trivial photoionisation delays.

\section{Results and discussions}
\subsection{Complex photoionisation time delay and the Wigner time delay}
The angle-differential and angle-integrated photoionisation
cross sections of single-photon ionisation (in the length gauge) can be expressed via the
corresponding dipole transition amplitudes as
\cite{hilborn1982einstein,natalense99a}
\begin{eqnarray}
\label{eq:sigma_def}
\sigma(E,\mathbf{\hat{k}}) = \frac{{\rm \pi}}{3c}
E{|D(E,\mathbf{\hat{k}})|}^2 
= \frac{\sigma(E)}{4{\rm \pi}}
\left( 1+\beta(E) P_2 \left( {\cos}(\mathbf{\hat{k}}\cdot \mathbf{\hat{e}}) \right) \right)
\\
    \sigma(E) = \frac{4{\rm \pi}^2}{3c}E{|D(E)|}^2 
\ .
\label{eq:sigma_int}
\end{eqnarray}
Here $E$ denotes the photon energy, $\mathbf{\hat{k}}$ and $\mathbf{\hat{e}}$
are the unit vectors pointing to the emission and polarisation
directions, respectively, and $c$ is the speed of
light  
\footnote{In the atomic units which are in use here and throughout
with  $e=m=\hbar=1$ and $c\approx 137$. The atomic unit of time
  1~a.u.~=~24.2~as.}.
The angular dependence in expression (\ref{eq:sigma_def}) enters via the
angular anisotropy parameter $\beta$ and the second Legendre
polynomial $P_2$.

\begin{figure}[t]
\includegraphics[width=\textwidth]{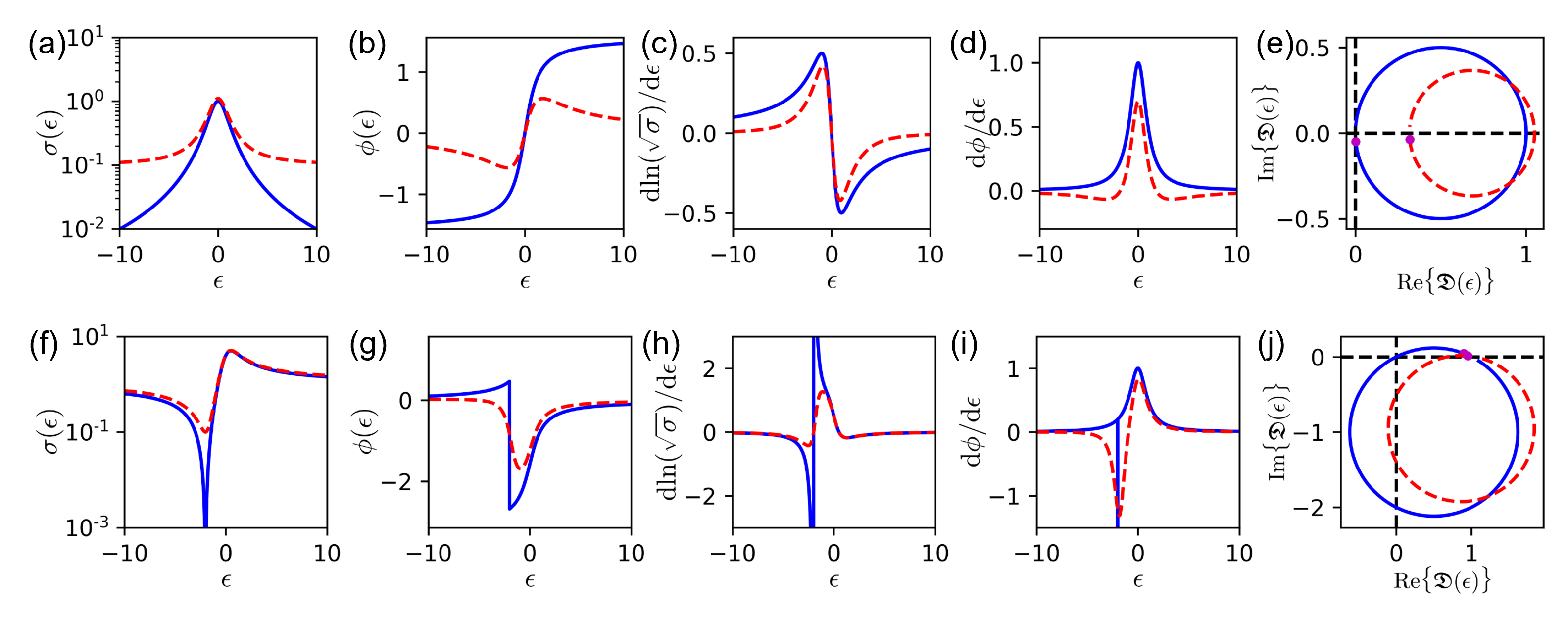}
\caption{\textbf{Transition amplitudes for the Lorentz and Fano
  resonances.} Cross sections (\textbf{a, f}), phases (\textbf{b, g}), the derivatives of the logarithm of the cross sections (\textbf{c, h}), Wigner time delays (\textbf{d, i}), and transition-amplitude trajectories
  (\textbf{e, j}) for the Lorentzian lineshape (\textbf{a-e}) and the Fano lineshape with $q = 2$ (\textbf{f-j}), respectively, where the blue solid lines correspond to $\sigma_{\rm a} = 1$ and $\sigma_{\rm b} \rightarrow 0^+$, while the red dashed lines correspond to $\sigma_{\rm a} = 1$ and $\sigma_{\rm b} = 0.1$. The magenta dots in (\textbf{e}) and (\textbf{j}) indicate the beginning of the trajectories at $\epsilon \rightarrow -\infty$. \label{fig:D_fano_lorentz} }
\end{figure}

The Wigner time delay
\cite{eisenbud1948formal,wigner1955lower,smith1960lifetime} is defined
as
\begin{eqnarray}
    \tau(E) = \frac{\partial {\rm arg} \{D(E)\}}{\partial E} 
    = \frac{\partial {\rm Im}\{\ln(D(E))\}}{\partial E} = {\rm Im}\left\{ \frac{\partial \ln(D(E))}{\partial E} \right\}. 
\label{eq:wigner_delay}
\end{eqnarray}
Photoionisation can be regarded as a half-scattering process
\cite{pazourek2015attosecond}, and its transition amplitude can be
presented by the $S$-matrix or the related $R$-matrix
\cite{wigner1947higher}. If the $S$-matrix is diagonalised with
respect to the angular momentum $\ell$ as
\begin{equation}
    S(k) = \sum_{\ell,m} \big| \ell,m 
\bigr\rangle {\rm e}^{2{\rm i}\phi_\ell} \bigl\langle \ell,m \big|
\label{eq:S_k}
\end{equation}
with $2\phi_\ell$ being the $\ell$-th scattering phase shift, then the
photoionisation amplitude corresponds to
\begin{eqnarray}
    S(k) - 1 = \sum_{\ell,m} \big| \ell,m 
    \bigr\rangle ({\rm e}^{2{\rm i}\phi_\ell} - 1) 
    \bigl\langle \ell,m \big| 
    = 2{\rm i} \sum_{\ell,m} \big| \ell,m 
    \bigr\rangle {\rm e}^{{\rm i}\phi_\ell} \sin\phi_l \bigl\langle \ell,m
    \big|
    \ .
\label{eq:S_k_minus1}
\end{eqnarray}
Here the eigenvalues have no longer moduli of 1, and the phases are
halved. Equation (\ref{eq:wigner_delay}) yields $\tau_\ell(E) = \partial
\phi_\ell / \partial E$, which is the original time delay proposed by Wigner
\cite{wigner1955lower}. The experimental time delay, on the other
hand, is usually expressed in the $\mathbf{\hat{k}}$-space with
interference of different $\ell$'s. Nonetheless,
equation (\ref{eq:wigner_delay}) can be interpreted as the ``generalised Wigner
delay'', which deals with the off-diagonal terms following Eisenbud's
formula
\cite{eisenbud1948formal,smith1960lifetime,kelkar2008analysis}.  The
analyticity of the $S$-matrix based on causality has been intensely
studied
\cite{schutzer1951connection,van1953s,gell1954use,goldberger1955use,karplus1955applications,khuri1957analyticity,meiman1964causality,taylor1972scattering}
and the scattering amplitude was shown to be analytical in the
upper-half complex $k$-plane for the physical region of the reaction,
where $k$ is the incident momentum. This is the foundation of the
complex-scaling method for practical computations
\cite{rescigno1973calculation,reinhardt1982complex,moiseyev1998quantum}. On
the other hand, the analyticity can also be expressed regarding
energy, and by applying the energy-time uncertainty principle, a
time-domain picture of the scattering processes arises
\cite{branson1964time,eden1965problem,peres1966causality}, where the
connection of this ``microscopic'' time and the attosecond time delay
is discussed in the next section. The analycity leads to the
Kramers-Kronig (KK) relations
\cite{kramers1924law,kronig1926theory,kramers1927diffusion} that allow
one to construct an energy-dependent complex function $f(E)$ using
only its real or imaginary part.  Previous works have used the
absorption cross section as the imaginary part to reconstruct the
frequency-domain response function of materials, where the time-domain
dynamics can be retrieved from their Fourier transform
\cite{abbamonte2004imaging,ott2013lorentz,stooss2018real}.  In this
work we show that using the analyticity of the ionisation amplitude,
${\rm Im}\{\ln(D(E))\}$ is similarly connected to its real part, and
the Wigner time delay can be retrieved via 
(\ref{eq:wigner_delay}).  The KK relation in the present form
requires that the absorption variation vanishes beyond a certain
frequency region, which is true for an (anti)resonance where the cross
section below and above the resonance region (in the vicinity of the
resonant energy $E_{\rm r}$) can be approximated as constant, and if
the variation of $E$ in equation (\ref{eq:sigma_def}) is negligible
compared to the drastic modulation of $D(E)$,
$\ln\left( \sigma(E) / \sigma_0 \right)$ can be used instead of $\ln\left((\sigma(E)/\sigma_0)/(E/E_0)\right)$, where $\sigma_0$ and $E_0$ are unitaries for the cross section and energy, respectively.  
The relation between the modulus of $D(E)$ and the cross section $\sigma(E)$ can be expressed as: 
\numparts
\begin{eqnarray}
    |D(E)| = \sqrt{\frac{3c}{4 {\rm \pi}^2 E} \sigma(E)} \\
    \fl \ln \left( |D(E)| \right) = {\rm Re} \left\{ \ln(D(E)) \right\} 
    = \frac{1}{2}\ln\left( \frac{\sigma(E) / \sigma_0}{E / E_0} \right) + {\rm const.} 
    \approx  \frac{1}{2}\ln\left( \frac{\sigma(E)}{\sigma_0} \right) + {\rm const.}
\label{eq:Re_ln_D_simple}
\end{eqnarray}
\endnumparts
The trajectory of $D(E)$ from $E \ll E_{\rm r}$ to $E \gg E_{\rm r}$
is approximately a closed contour on the complex plane, as shown in
figure \ref{fig:D_fano_lorentz} (e) and (j). Using the KK relations of
the amplitude and the phase of a complex function, it has been shown
\cite{lee1932synthesis,bode1945network,raymond1951transformees,hoenders1975solution,burge1976phase,mecozzi2009retrieving,mecozzi2016necessary}
that the phase can be retrieved (with a freedom of a common phase shift)
from the absolute value when $\ln(D(E))$ vanishes faster than
$E^{-1}$ and $D(E)$ fulfils the minimum-phase condition, namely,
that all the poles and zeros are located in the same half-plane. This
means that the origin is not enclosed in the trajectory of $D(E)$ (the
winding number is 0), and the total phase change throughout the
(anti)resonant region is 0 instead of $\pm 2{\rm \pi}$
\cite{burge1976phase,mecozzi2016necessary}.

In the interaction picture $\hat{H}(\Delta \mathfrak{t}) = \hat{H}_0 +
\hat{H}_1(\Delta \mathfrak{t})$, where $\hat{H}_0$ is the field-free
Hamiltonian, and $\hat{H}_1(\Delta \mathfrak{t})$ is the interaction
term as a function of retardation, the interaction is turned on at
$\Delta \mathfrak{t} = 0$, which triggers a transition at $\Delta
\mathfrak{t} \ge 0$. This retardation is the ``microscopic'' time
introduced by Branson \cite{branson1964time}, and we have ignored the
relativistic effect of the virtual-particle ``cloud'', whose temporal
feature is much shorter than the electronic response in atoms
(molecules), as analogously shown in a previous study \cite{peres1966causality}. Thus,
in the energy (frequency) domain, we have
\begin{eqnarray}
    D(E) = \int_{-\infty}^{+\infty} \tilde{D}(\Delta \mathfrak{t}) {\rm e}^{{\rm i}E \Delta \mathfrak{t}} {\rm d}{\Delta \mathfrak{t}} 
    = \int_{0}^{+\infty} \tilde{D}(\Delta \mathfrak{t}) {\rm e}^{{\rm i}E \Delta \mathfrak{t}} {\rm d}{\Delta \mathfrak{t}}
\label{eq:D_E_reciprocal}
\end{eqnarray}
where $\tilde{D}(\Delta \mathfrak{t})$ has the form: 
\begin{eqnarray}
    \tilde{D}(\Delta \mathfrak{t}) =
    \cases{\frac{1}{2 \pi} \int_{-\infty}^{+\infty} D(E) {\rm e}^{-{\rm i}E \Delta \mathfrak{t}} {\rm d}E 
    &for $ \Delta \mathfrak{t} \ge 0 $ \\
    0 &for $ \Delta \mathfrak{t} < 0 $}.
\label{eq:D_E_invFourier}
\end{eqnarray}
The case of $\Delta \mathfrak{t} \geq 0$ corresponds to the time propagation of a superposition state, while the elimination of the transition amplitude for $\Delta \mathfrak{t} < 0$ reflects the causal condition. 
{{If we perform analytical continuation of $E = \mathcal{E}+{\rm i}\mathfrak{E}$, where $\mathcal{E}$ and $\mathfrak{E}$ are real and $\mathfrak{E} > 0$, equation (\ref{eq:D_E_reciprocal}) becomes:}}  
\begin{equation}
    D(E) = D(\mathcal{E}+{\rm i}\mathfrak{E}) = \int_0^{+\infty} \tilde{D}(\Delta \mathfrak{t}) {\rm e}^{{\rm i} \mathcal{E} \Delta \mathfrak{t}} {\rm e}^{-\mathfrak{E} \Delta \mathfrak{t}} {\rm d}{\Delta \mathfrak{t}}.
\label{eq:D_E_reciprocal_complex}
\end{equation}
Since the dipole transition amplitude of real energy $D(\mathcal{E}) = \int_0^{+\infty} \tilde{D}(\Delta \mathfrak{t}) {\rm e}^{{\rm i} \mathcal{E} \Delta \mathfrak{t}} {\rm d}{\Delta \mathfrak{t}}$ converges and ${\rm e}^{-\mathfrak{E} \Delta \mathfrak{t}}$ is a bounded and monotonically decreasing real function for $\Delta \mathfrak{t} \geq 0$, equation (\ref{eq:D_E_reciprocal_complex}) fulfils the Abel-Dirichlet test of convergence for improper integrals and thus converges \cite{shilov1996elementary,Zorich2016}. This ensures that $D(E)$ has no poles in the upper-half of the complex energy plane
\footnote{Unlike $D(E)$, any pole of $\sigma(E) \propto {|D(E)|}^2$ will have a complex conjugate pole on the other half plane, since ${|D(E)|}^2 = D(E) D^*(E)$. Both $D(E)$ and $D^*(E)$ correspond to the same $\sigma(E)$, but $D^*(E)$ has the opposite phase, which corresponds to the reversed causality.}.

From
expression (\ref{eq:D_E_reciprocal}), we can define the complex ``average
retardation'', with the same expression derived by Pollak and Miller
\cite{pollak1984new,yamada2004unified}, as
\begin{eqnarray}
    \fl \mathfrak{T}(E) = \frac{\int_0^{+\infty} {\Delta \mathfrak{t}}
      \tilde{D}(\Delta \mathfrak{t}) {\rm e}^{{\rm i}E \Delta
        \mathfrak{t}} {\rm d}{\Delta \mathfrak{t}}}{\int_0^{+\infty}
      \tilde{D}(\Delta \mathfrak{t}) {\rm e}^{{\rm i}E \Delta
        \mathfrak{t}} {\rm d}{\Delta \mathfrak{t}}} 
    =
    \frac{1}{\rm i} \frac{1}{D(E)} \frac{\partial D(E)}{\partial E} =
    \frac{1}{\rm i} \frac{\partial \ln(D(E))}{\partial E} ~ .
\end{eqnarray}
If we assume that the minimum-phase condition is fulfilled (we shall come
back to this point in the discussion of the CM), the
Wigner time delay $\tau(E)$ is connected to $\mathfrak{T}(E)$ and
$\sigma(E)$ using equations (\ref{eq:wigner_delay}) and
(\ref{eq:Re_ln_D_simple}) and the KK relations by applying the
logarithm Hilbert transform (LHT):
\begin{eqnarray}
    \fl \tau(E) = {\rm Im}\{{\rm i} \mathfrak{T}(E) \} = {\rm Re}\{ \mathfrak{T}(E) \} = -\mathcal{H}\{ {\rm Im}\{ \mathfrak{T}(E) \} \} \nonumber\\
    =\frac{1}{2} \frac{\partial}{\partial E} \mathcal{H}\{ \ln(\sigma(E)) \} = \frac{1}{2} \mathcal{H} \left\{ \frac{1}{\sigma(E)} \frac{\partial \sigma}{\partial E} \right\}
\label{eq:tau_Hilbert}
\end{eqnarray}
where $\mathcal{H}$ is the Hilbert transform:
\begin{equation}
    \mathcal{H}\{f(E)\} = \frac{1}{{\rm \pi}E} \otimes f(E) 
    = \frac{1}{{\rm \pi}} \mathcal{P} \int_{-\infty}^{+\infty} \frac{f(E^\prime)}{E - E^\prime} {\rm d}{E^\prime}
\label{eq:def_Hilbert}
\end{equation}
which is linear and commutative to the derivative. Here $\otimes$
represents the convolution, and $\mathcal{P}$ refers to the Cauchy
principal value. 

Relation (\ref{eq:tau_Hilbert}) establishes the general method to retrieve the Wigner time delay from the photoionisation cross section. However, as the integral (\ref{eq:def_Hilbert}) runs from $-\infty$ to $+\infty$, including the bound states for the negative part, using the whole profile of the photoionisation cross section is not practical. Instead, one can investigate the behaviour near a resonance with energy $E_{\rm r}$ above the threshold, for instance, the shape resonance, where the variation of the transition amplitude rapidly decays beyond a relatively small energetic window in the vicinity of $E_{\rm r}$, and the Wigner time delay shows thus local behaviour within this region. 
The boundary conditions at (positive and negative) infinity hence become
\begin{equation}
    \lim_{(E - E_{\rm r}) \rightarrow \infty} \frac{|E - E_{\rm r}|}{\sigma(E)} \frac{\partial \sigma}{\partial E} = 0 ~ .
\label{eq:boundary}
\end{equation}
For example, for the Lorentzian profile, the cross-section variation ${\partial \sigma}/{\partial E}$ is proportional to ${|E - E_{\rm}|}^{-3}$. If we assume that $\sigma(E)$ outside the resonant region is approximately constant due to the contribution of the non-resonant channels, expression (\ref{eq:boundary}) is fulfilled. In the following sections, we will give the analytical expressions of the Wigner time delay for the Lorentzian and Fano lineshapes using LHT. We will demonstrate that LHT is also applicable to resonances with multi-peak features, for example, the xenon giant resonance modulated by a C$_{60}$ cage. Finally, we will examine the Cooper minima and discuss the role of the minimum-phase condition, i.e. the winding number of the trajectory. 

\subsection{(Anti)resonances with Lorentzian and Fano lineshapes}
The cross section of an (anti)resonance can be expressed by the
Lorentzian lineshape, e.g., the Breit-Wigner resonance formula
\cite{breit1936capture,friedrich2017theoretical}:
\begin{equation}
    \sigma_{\rm L}(\epsilon) = \sigma_{\rm a} \frac{1}{\epsilon^2+1} + \sigma_{\rm b}
\label{eq:lorentz_peak}
\end{equation}
or the Fano lineshape \cite{fano1961effects,fano1965line}:
\begin{equation}
    \sigma_{\rm F}(\epsilon) = \sigma_{\rm a} \frac{{(\epsilon+q)}^2}{\epsilon^2+1} + \sigma_{\rm b}
\label{eq:fano_peak}
\end{equation}
where $\epsilon = (E-E_{\rm r})/(\Gamma/2)$ is the relative energy,
and $\sigma_{\rm a}$ and $\sigma_{\rm b}$ are the resonant and
non-resonant cross sections, respectively. We assume that the
corresponding pathways contribute to the overall transition amplitude
coherently, which fits in the Feshbach picture
\cite{feshbach1962unified} and was recently demonstrated in the X-ray
regime \cite{ma2022first}. The Lorentzian lineshape is symmetric,
while the Fano lineshape is asymmetric, with $q$ being the asymmetric
parameter. $\Gamma$ describes the peak width and thus
the resonant state has a lifetime of $1/\Gamma$ from Fano's picture. The cross section for a Lorentzian or Fano lineshape can be written as a common expression:  
\begin{equation}
    \sigma_{{\rm L/F}}(\epsilon) = \mathfrak{S}
    \frac{(\epsilon+Q)^2+\gamma^2}{\epsilon^2+1} ~ .
    \label{eq:sigma_general}
\end{equation}
For the Lorentzian lineshape, $Q = 0$, $\gamma^2 = 1 + \sigma_{\rm a}/\sigma_{\rm b}$, where $\mathfrak{S} = \sigma_{\rm b}$, 
while for the Fano lineshape, $Q =
q/(r+1)$, $\gamma^2 = r(r+q^2+1)/(r+1)^2$, where $r = {\sigma_{\rm b}/\sigma_{\rm a}}$, and
$\mathfrak{S} = \sigma_{\rm a} + \sigma_{\rm b}$. The detailed proof
can be found in \ref{sec:appendix_lorentz_fano}. Using $\partial/\partial E =
(2/\Gamma)\partial/\partial \epsilon$, we have
\begin{equation}
    \frac{1}{\sigma(E)} \frac{\partial \sigma}{\partial E}
    = \frac{2}{\Gamma} \left[ \frac{\frac{2}{\gamma} \left(\frac{\epsilon+Q}{\gamma}\right)}{{\left(\frac{\epsilon+Q}{\gamma}\right)}^2+1} 
    - \frac{2\epsilon}{\epsilon^2+1} \right].
\label{eq:dsigma_dE_sigma}
\end{equation}
It is easy to verify that the boundary conditions
(\ref{eq:boundary}) are fulfilled, and (\ref{eq:tau_Hilbert})
yields
\begin{equation}
    \tau_{{\rm L/F}}(E) = \frac{2}{\Gamma} \left[
      -\frac{\frac{1}{|\gamma|}}{{\left(\frac{\epsilon+Q}{|\gamma|}\right)}^2+1}
      + \frac{1}{\epsilon^2+1} \right]
\ .
\label{eq:tau_analytical}
\end{equation}
Here we used the property of the Hilbert transform that
$\mathcal{H}\{f(aE)\}(E) = {\rm sgn}(a)\times\mathcal{H}\{f(E)\}(aE)$. 
The time delay can be decomposed as two Lorentzian peaks with widths of $\gamma \Gamma$ and $\Gamma$ and opposite signs, centred at $(E_{\rm r}-\frac{Q\Gamma}{2})$ and $E_{\rm r}$, respectively. 
For the Fano lineshape, the positive and negative peaks are separated by $Q \Gamma/2$. When $r \rightarrow 0$ and thus $|\gamma| \rightarrow 0$, the first time-delay peak approaches $-{\rm \pi}\delta(\epsilon+Q)$. These correspond to the cases plotted in figure \ref{fig:D_fano_lorentz} (i).
On the other hand, for the Lorentzian lineshape, since $Q=0$, the two peaks are co-centred and their sum is symmetric. If $\sigma_{\rm b}/\sigma_{\rm a} \rightarrow 0$ and thus $|\gamma| \rightarrow \infty$, the first peak vanishes and thus the time delay is a single Lorentzian peak with $\tau(E_{\rm r}) = 2/\Gamma$ at its centre, as discussed in \cite{goldberger1962concerning}. For the general case, it corresponds to a narrower peak subtracted by a wider peak, yielding the alternating sign of the time delay. These cases are illustrated in figure \ref{fig:D_fano_lorentz} (d).

Since $(Q + {\rm i}\gamma)$ can be regarded as the ``complex
asymmetry parameter'' \cite{ma2022first}, a more insightful approach
is to investigate an analytical function that satisfies ${|\mathfrak{D}(\epsilon)|}^2 = \sigma(\epsilon)$, as proposed in
\cite{kotur2016spectral,argenti2017control}: 
\begin{equation}
    \mathfrak{D}(\epsilon) = \sqrt{\mathfrak{S}} \frac{\epsilon+Q+{\rm i}\gamma}{\epsilon+{\rm i}} = \sqrt{\mathfrak{S}} \left[ 1 + \frac{Q + {\rm i}(\gamma - 1)}{\epsilon + {\rm i}} \right].
\label{eq:D_epsilon}
\end{equation}
It fulfils the minimum-phase condition when $\gamma > 0$, so $\mathfrak{D}(\epsilon)$ is related to $D(\epsilon)$ by multiplying $\sqrt{\frac{4 \pi^2 E}{3c}}$ and a common phase offset, namely $D(\epsilon) = \sqrt{\frac{4 \pi^2 E}{3c}} \mathfrak{D}(\epsilon) {\rm e}^{{\rm i}\phi_0}$, where $\sqrt{\frac{4 \pi^2 E}{3c}}$ is approximately constant in the vicinity of the (anti)resonance, and
\begin{eqnarray}
\fl \arg{\{D(\epsilon)\}} = \arg{\{\mathfrak{D}(\epsilon)\}} + \phi_0
  = \arg{\{\epsilon+Q+{\rm i}\gamma\}} - \arg{\{\epsilon+{\rm i}\}} + \phi_0 \nonumber\\
  = \cot^{-1}\left(\frac{\epsilon+Q}{\gamma}\right) -
  \cot^{-1}(\epsilon) + \phi_0 ~ .
\label{eq:D_mathfrakD}
\end{eqnarray}
%
Therefore, $D(\epsilon)$ and $\mathfrak{D}(\epsilon)$ share the same Wigner time delay $\tau(\epsilon) = \frac{2}{\Gamma} {\partial {\rm arg} \{D(\epsilon)\}}/{\partial \epsilon}$, which is expressed by (\ref{eq:tau_analytical}).
The trajectory of $\mathfrak{D}(\epsilon)$
from $\epsilon \rightarrow -\infty$ to $\epsilon \rightarrow +\infty$
is a counter-clockwise rotating (due to the causality) closed circle
\cite{taylor1972scattering,kotur2016spectral} with the radius of
$\frac{\sqrt{\mathfrak{S}}}{2}\sqrt{{(\gamma - 1)}^2 + Q^2}$ centred
at $\left( \sqrt{\mathfrak{S}}\frac{\gamma+1}{2},
-\sqrt{\mathfrak{S}}\frac{Q}{2} \right)$. It can be easily verified
that the origin is not enclosed when $\gamma > 0$. Take the Fano
resonance as an example, at $\epsilon \rightarrow \pm \infty$, the
transition amplitudes are dominated by the direct ionisation channel,
which is approximately constant in this region. As $r \rightarrow 0$
(``pure'' Fano lineshape), the trajectory approaches tangential to the
origin, where a $(-{\rm \pi})$-phase jump occurs within an infinitely
small energy interval, which corresponds to the $-{\rm \pi}\delta(\epsilon+Q)$ peak.
{As $\gamma$ deviates from 0, the maximal phase jump is $ 2 {\rm sgn}(\gamma-1) {\cos}^{-1}\left( 2\sqrt{\gamma} / \sqrt{{(\gamma + 1)}^2 + Q^2} \right)$ for $\gamma \neq 1$ and $\pm 2 {\tan}^{-1}(Q/2)$ for $\gamma=1$ (see \ref{sec:appendix_phase_jump} for details).} 
For the ``pure'' Lorentzian transition
amplitude $D(\epsilon) \propto (\epsilon+{\rm i})^{-1}$
\cite{taylor1972scattering}, there is a phase offset of ${\rm \pi}$
between $\epsilon \rightarrow +\infty$ and $\epsilon \rightarrow
-\infty$, which does not fulfil the minimum-phase condition since it
encloses ``half'' the origin. Its phase retrieved by the LHT method
starts and ends at the same value at infinities, which corresponds to
shifting the circle by an infinitely small amount away from the
origin, and thus subtracts an infinitely broad Lorentzian lineshape
according to (\ref{eq:tau_analytical}), where $Q = 0$ and $\gamma
\rightarrow +\infty$. Its behavior near $\epsilon = 0$, which is of
most interest, is largely unchanged. 
{The maximal phase jump is $4 {\tan}^{-1}(\sqrt{\gamma}) - \pi = 4 {\tan}^{-1}(\sqrt[4]{1+\sigma_{\rm a}/\sigma_{\rm b}}) - \pi$, which approaches $\pi$ when $\sigma_{\rm a}/\sigma_{\rm b} \rightarrow 0$.}
Examples of Lorentzian and Fano
lineshapes are plotted in figure \ref{fig:D_fano_lorentz}, where the
increase of $\sigma_{\rm b}/\sigma_{\rm a}$ results in the decrease of the extreme(s) of the time
delay.  

\begin{figure}[bt]
\includegraphics[width=\textwidth]{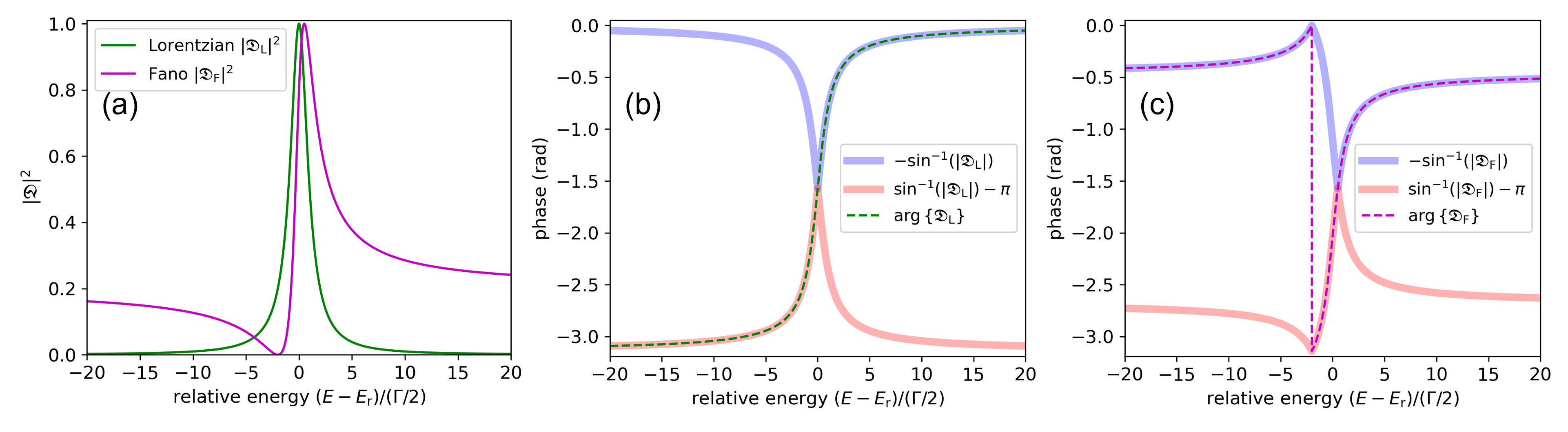}
\caption{\textbf{Phase retrieval for Lorentzian and Fano lineshapes using $\sigma_l(E)\propto \sin^2\phi_l$.} Normalized cross section of the Lorentzian and Fano lineshapes with $\sigma_{\rm b}=0$ are plotted in (\textbf{a}). For the Lorentzian lineshape $\mathfrak{D}_{\rm L}(\epsilon) = 1/(\epsilon + {\rm i})$, while for the Fano lineshape $\mathfrak{D}_{\rm F}(\epsilon) = {(\epsilon+q)}/{\left( (q+{\rm i})(\epsilon+{\rm i}) \right)}$, where $q=2$ is used for plotting. The phase of $\mathfrak{D}(\epsilon)$ and the phase retrieved according to ($\phi(\epsilon) = \sin^{-1}(|\mathfrak{D}(\epsilon)|)$) for the Lorentzian and Fano lineshapes are shown in (\textbf{b}) and (\textbf{c}), respectively. Note that adding the phase by a constant factor will not change the corresponding Wigner time delay defined in equation (\ref{eq:wigner_delay}). \label{fig:sin2_fano_lorentz}}
\end{figure}

In preceding work \cite{kheifets2023shape} the shape-resonance
analysis was performed by utilising the relation $\sigma_l(E)
\propto \sin^2\phi_l$ from equation (\ref{eq:S_k_minus1}), which is
applicable if only one partial wave of angular momentum $l$ is dominating.
{Assume that no background is contributing, and let $f_l(E) = -\cot \left( \phi_l(E) \right) $ be a real function, then}
\begin{equation}
    \sin^2 \left( \phi_l(E) \right) = \frac{1}{{f_l(E)}^2+1} = {\Big| \frac{1}{f_l(E) \pm {\rm i}} \Big|}^2 ~ .
    \label{eq:CS_sin2_Lorentz}
\end{equation}
{Therefore, assuming $f_l(E)$ to be analytical,}
\begin{eqnarray}
    \mathcal{H} \Bigl\{ \frac{1}{2}\ln \left( \sigma_l(E) \right) \Bigr\} = -\arg\{ f_l(E)\pm{\rm i} \} 
    = \mp \cot^{-1} \left( f_l(E) \right) = \pm \phi_l(E)
    \label{eq:Hilbert_sin2_Lorentz}
\end{eqnarray}
if one of the two branches $(f_l(E)\pm{\rm i})$ has no poles or zeros on the upper half-plane of $E$. For example, let $f_l(E) = \epsilon$, where $\epsilon = (E - E_{{\rm r}, l})/(\frac{\Gamma_l}{2})$ and $\Gamma_l>0$, $(f_l(E)+{\rm i})$ fulfils the condition and leads to the ``pure'' Breit-Wigner resonance with Lorentzian lineshape. For the ``pure'' Fano lineshape we have $f_l(E) = (q \epsilon - 1) / (\epsilon + q)$ thus $f_l(E) + {\rm i} = (q + {\rm i})(\epsilon + {\rm i}) / (\epsilon + q)$ also fulfils the requirement. Since $\pm \phi_l(E)$ and $\pi \pm \phi_l(E)$ yield the same value in expression (\ref{eq:CS_sin2_Lorentz}), the Wigner time delay corresponds to $\pm \partial \arg\{ \phi_l(E) \} / \partial E$, and the sign depends on different regions, as shown in figure \ref{fig:sin2_fano_lorentz}.

\begin{figure}[bt]
\includegraphics[width=\textwidth]{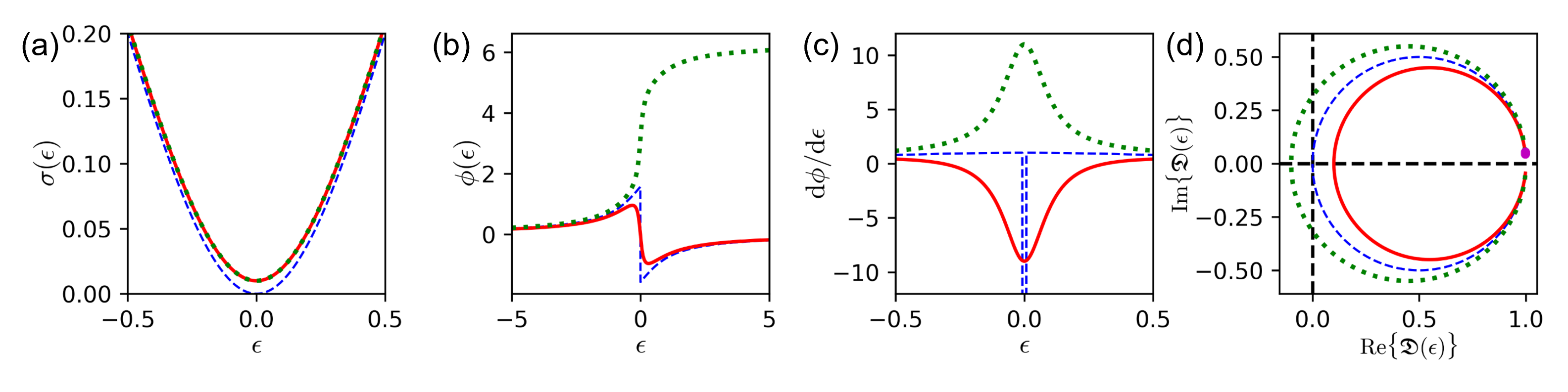}
\caption{\textbf{Comparison of the trajectories with different topologies.} Cross sections in the vicinity of the minimum (\textbf{a}), phases (\textbf{b}), time delays (\textbf{c}), and trajectories (\textbf{d}, the magenta dot indicates $\epsilon \rightarrow -\infty$) of the anti-Lorentzian lineshapes with $\mathfrak{S} = 1$, $Q = 0$ and $\gamma = 0^{+}$ (blue, dashed), $0.1$ (red, solid), or $-0.1$ (green, dotted), according to (\ref{eq:D_epsilon}). \label{fig:D_Cooper} }
\end{figure}

In real experiments, the measurable energy range is finite and thus the boundary conditions are not strictly fulfilled. This leads to a
deviation of the retrieved time delay from its physical
counterpart. However, when comparing two photoionisation processes,
the relative time delay, which is usually the experimental
observable \cite{kluender11a}, can be obtained from their
cross-section ratio, which converges faster than the individual time
delays:
\begin{eqnarray}
    \fl \Delta \tau(E) = \tau_1(E) - \tau_2(E) = \frac{\partial}{\partial E} \arg \left\{\frac{D_1}{D_2}\right\} \nonumber\\
    = \frac{1}{2} \frac{\partial}{\partial E} \mathcal{H} \left\{ \ln \left(\frac{\sigma_1}{\sigma_2}\right) \right\} = 
    \frac{1}{2} \mathcal{H} \left\{ \frac{\sigma_2}{\sigma_1} \frac{\partial(\sigma_1/\sigma_2)}{\partial E} \right\} ~ .
\end{eqnarray}
This is particularly useful for comparing the same resonance under different surroundings, for example the giant resonance of xenon with and without a C$_{60}$ cage, as discussed in the next section. 

\subsection{Composite resonances with more complicated features}
\begin{figure} 
\begin{center}
\includegraphics[width=0.8\textwidth]{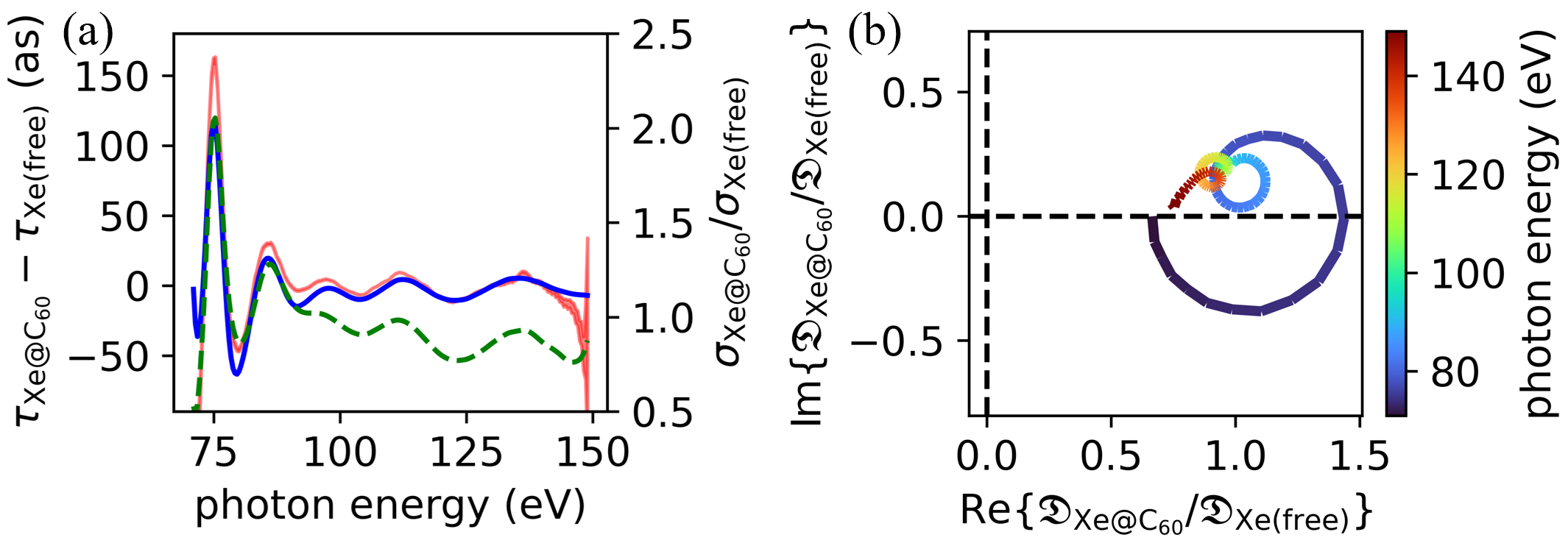}
\end{center}
\caption{Applications of the LHT formalism to Xe and Xe@C$_{60}$. (a) The calculated cross-section ratio (green dashed, right vertical axis) using the RPAE method and the time delay (blue, left vertical axis) \cite{bray2018photoionization} and the time delay retrieved from the cross sections using the LHT (red, left vertical axis). The filled area indicates the difference between first taking the energy derivative then perform the Hilbert transform and first performing the Hilbert transform then taking the energy derivative, which are the same when the boundary conditions are strictly fulfilled. (b) The transition-amplitude trajectory constructed by $\mathfrak{D}_{\rm Xe@C_{60}} / \mathfrak{D}_{\rm Xe(free)} = \exp[\frac{1}{2} \ln(\mathfrak{r}) + \frac{\rm i}{2} \mathcal{H}\{ \ln(\mathfrak{r}) \}]$, where $\mathfrak{r} = {\sigma_{\rm Xe@C_{60}}}/{\sigma_{\rm Xe(free)}}$ is the cross-section ratio. \label{fig:Xe_C60_rel}}
\end{figure}

We note that the amplitude-phase relation can be extended beyond ``simple'' resonances with Lorentzian or Fano lineshapes, as the derivation from the causality is more general than the specific models of transition amplitude. For example, the cross sections and time delays of free xenon and xenon in a ${\rm C}_{60}$ cage in the region of the giant resonance have been computed using the random-phase approximation (RPA), relativistic RPA (RRPA) \cite{deshmukh2014attosecond}, RPA with exchange (RPAE), and time-dependent Schr{\"o}dinger equation (TDSE) methods \cite{bray2018photoionization}. As shown in figure \ref{fig:Xe_C60_rel}, the cross-section ratio shows multiple peaks in the giant-resonance region, which reflects the influence of the ${\rm C}_{60}$ shell, and the time delay retrieved by the LHT, which does not require the phase of the transition amplitude, almost quantitatively agrees with the reported value obtained from the phase. Positive (negative) time delays are typically associated with local maxima (minima) of the cross-section ratio. The reconstructed trajectory of ${\mathfrak{D}_{\rm Xe@C_{60}}}/{\mathfrak{D}_{\rm Xe(free)}}$ has a multiple-spiral structure, which manifests the validity of the LHT method for complex resonances. This relies on the fact that the modulation of the ${\rm C}_{60}$ surroundings is smaller than the modulus of the transition amplitude $per~se$, so that the transition-amplitude ratio lies around $(1,0)$ on the complex plane, as illustrated in figure \ref{fig:Xe_C60_rel} (b), which avoids encircling the origin, and its winding number can thus be taken to be zero.

\subsection{Time delays near Cooper minima}
The Cooper minima are antiresonances that have a distinctly different origin compared to the Fano or shape resonances \cite{cooper62a,woerner09a,schoun14a}. Nevertheless, the cross section near the minimum can be locally fitted by a Fano lineshape with $\gamma \rightarrow 0$ in expression (\ref{eq:sigma_general}) (see \ref{sec:appendix_Fano_CM} for details). Although $\pm \gamma$ give the same cross section, they have different topological structures on the complex plane. Only the transition amplitude corresponding to the positive $\gamma$ is minimum-phase, which leads to a negative time delay and is expressed by (\ref{eq:tau_analytical}) using the same fitting parameters, while the negative $\gamma$ gives a positive time delay, as demonstrated in figure \ref{fig:D_Cooper}. We can quantify the relative shift of the trajectory by defining the parameter: 
\begin{equation}
    g = \frac{|\gamma|}{Q^2 + 1} \approx \sqrt{\frac{r}{q^2 + 1}}
\end{equation}
which is the distance between the origins of the circles corresponding to $\pm \gamma$, divided by the average diameter, when $r \approx 0$ and thus $|\gamma| \ll 1$. For example, in figure \ref{fig:D_Cooper} (d), $g = 0.1$, where $r \approx 0.01$ and $q = 0$.

For numerical illustration of the proposed technique, we analyse the
time delay near the CM in various shells of noble gas
atoms: Xe $4d$, Ar $3p$ and Ar $3s$. The Xe $4d$ and Ar $3p$ CM have a kinematic origin where the radial node in the target
orbital passes through the oscillation in the continuum radial
orbital. The case of Ar $3s$ is different as the corresponding CM is induced by the inter-shell correlation with the $3p$
shell. These two different cases of kinematic and correlation induced
CM allows us to demonstrate the utility of the proposed
technique. Our numerical illustrations are based on the RPAE
calculations reported in \cite{kheifets2013time}, and more information can be found in \ref{sec:appendix_RPAE}. First we evaluate
the angle differential photoionisation cross section
(\ref{eq:sigma_def}) in the zero emission direction $\mathbf{\hat
  k}\parallel\mathbf{\hat e}$. This cross section in the vicinity of the
CM is fitted with the standard set of Fano parameters
  $E_{\rm r}$, $\Gamma$, $q$ and  $\rho^2$ \cite{fano1961effects,fano1965line,kossmann1988new}. Next,
these parameters are converted to an alternative set $Q$ and $\gamma$
which allows to express the photoionisation amplitude
${\mathfrak{D}}(E)$ (\ref{eq:D_epsilon}, \ref{eq:D_mathfrakD}) and the corresponding
time delay $\tau(E)$ (\ref{eq:tau_analytical}). These two
quantities are compared with their numerical counterparts evaluated using
the RPAE method. 

\begin{figure}
\begin{center}
\includegraphics[width=0.90\textwidth]{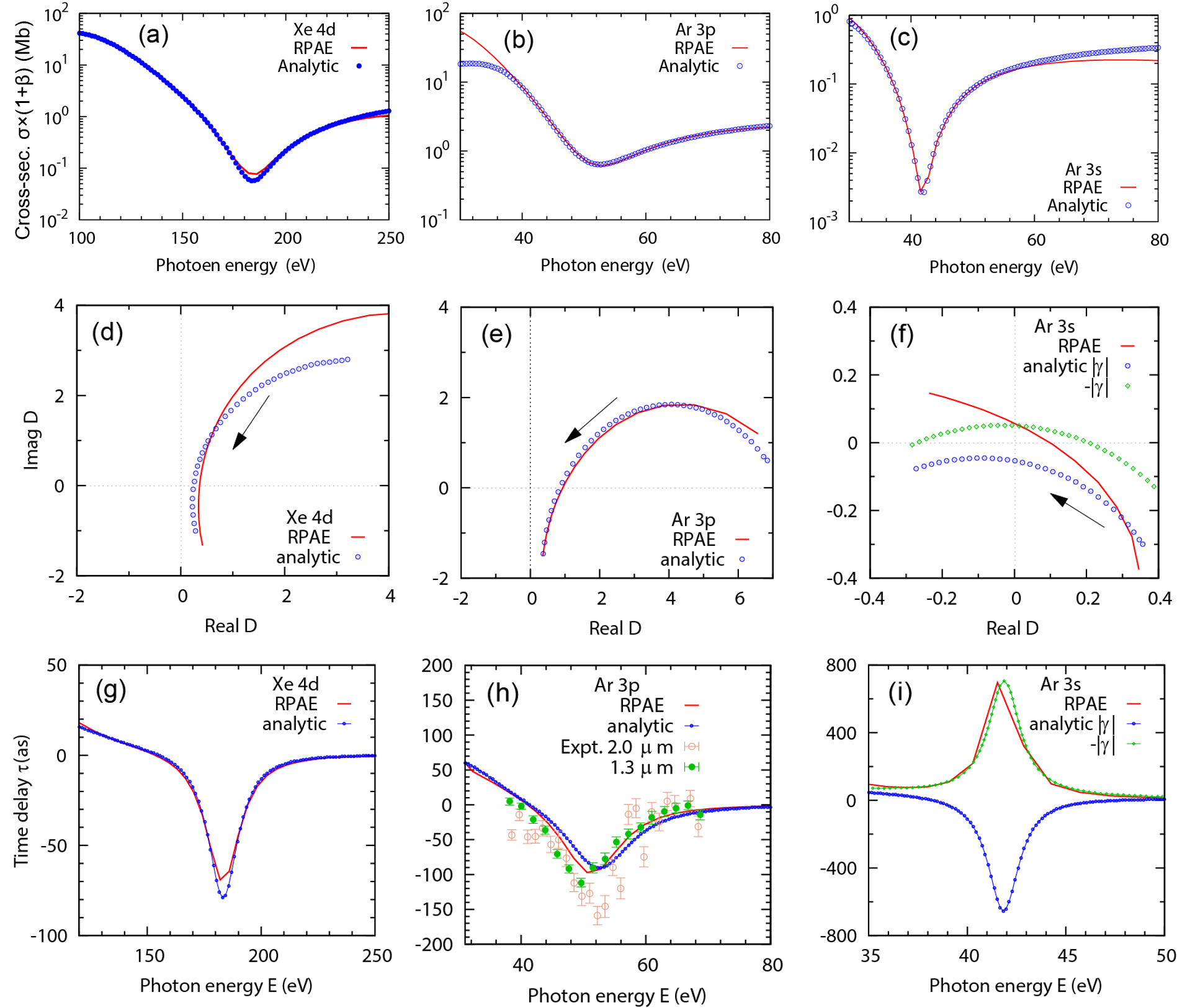}
\end{center}
\caption{\textbf{Comparison of the cross-section-time-delay relation
    of various Cooper minima.} The photoionization cross sections in
  the Cooper minima of Xe $4d$ (\textbf{a}), Ar $3p$ (\textbf{b}), and
  Ar $3s$ (\textbf{c}) from the RPAE calculations (solid red) are
  fitted with the Fano lineshape (\ref{eq:fano_peak}) (blue dots).  The
  photoionisation amplitudes of Xe $4d$ (\textbf{d}), Ar $3p$
  (\textbf{e}), and Ar $3s$ (\textbf{f}) computed by RPAE (red) and
  reconstructed from the Fano profile using (\ref{eq:D_epsilon})
  (blue) are compared. The corresponding time delays from RPAE and by
  the analytical formula in (\ref{eq:wigner_delay}) are shown in
  (\textbf{g-i}). \label{fig:Cooper}}
\end{figure}

This comparison is displayed in figure \ref{fig:Cooper}. Panels (a-c)
show the photoionisation cross sections near the corresponding CM. 
In panels (d-f), the parametric plots exhibit the complex
photoionisation amplitudes $\mathfrak{D}(E)$ as given analytically by
(\ref{eq:D_epsilon}) and evaluated in the RPAE, respectively.  The
arrows in the amplitude graphs indicate the winding direction as the
photon energy increases. In (g-i), we display the
corresponding time delays. The three columns, from left to right,
correspond to the Xe $4d$, Ar $3p$ and Ar $3s$ orbitals, respectively. 
{The LHT-extracted time delays given by (\ref{eq:fano_peak}) show very good agreement with the numeric results for Xe $4d$ and Ar $3p$, while for the Ar $3s$ CM there is a qualitative disagreement that LHT yields negative delay at the cross-section minimum, while the computation suggests the opposite.}
Such a profound difference in time delays near the CM can
be traced to the corresponding photoionisation amplitudes and
their winding numbers, as plotted in panels (d-f). In the cases of Xe $4d$ and Ar $3p$, both the
analytical and numeric amplitudes do not encircle the origin and their
respective winding numbers are 0. The case of Ar $3s$ is different
where the RPAE amplitude encircles the origin, namely, the minimum-phase condition is not fulfilled.
{In equation (\ref{eq:D_epsilon}), when $\gamma < 0$, the transition amplitude, denoted as $\breve{D}(\epsilon)$, has a zero at $(-Q+{\rm i}|\gamma|)$ and thus has winding number of 1. The LHT-retrieved transition amplitude can be expressed as}
\begin{equation}
    \mathfrak{D}(\epsilon) = \frac{\epsilon+Q+{\rm i}|\gamma|}{\epsilon+Q-{\rm i}|\gamma|} \sqrt{\frac{3c}{4 \pi^2 E}} \breve{D}(\epsilon) {\rm e}^{-{\rm i} \phi_0}
\label{eq:D_winding_1}
\end{equation}
{which gives rise to a Lorentzian lineshape that should be added to the LHT-retrieved time delay:}
\begin{equation}
    \frac{\partial \arg\{ \breve{D}(\epsilon) \}}{\partial \epsilon}
    = \frac{\partial \arg\{ \mathfrak{D}(\epsilon) \}}{\partial \epsilon}
    + \frac{\frac{2}{|\gamma|}}{{\left( \frac{\epsilon+Q}{|\gamma|} \right)}^2 + 1} ~ .
\label{eq:dphi_depsilon_winding_1}
\end{equation}
{Hence, (\ref{eq:tau_analytical}) becomes}
\begin{equation}
    \breve{\tau}(E) = \frac{2}{\Gamma} \left[\frac{\frac{1}{|\gamma|}}{{\left(\frac{\epsilon+Q}{|\gamma|}\right)}^2+1} + \frac{1}{\epsilon^2+1} \right]
\label{eq:tau_analytical_winding_1}
\end{equation}
{which turns the negative peak into a positive peak, and in the limit of $\gamma \rightarrow 0^{-}$, it becomes the $+{\rm \pi}\delta(\epsilon+Q)$ peak.
The time delays near the CM for the zero-winding-number and one-winding-number trajectories are given by (\ref{eq:tau_analytical}) and (\ref{eq:tau_analytical_winding_1}), respectively.}
Using equations (\ref{eq:D_winding_1}, \ref{eq:tau_analytical_winding_1}), the complex transition amplitude and the corresponding time delay of the Ar $3s$ CM with winding number equalling to 1 can be retrieved, as shown in panels (f) and (i) in figure \ref{fig:Cooper}, respectively. This gives a positive time delay that excellently agrees with the RPAE result, which indicates that the KK relations reduces the detailed time-delay computation into the quantitative winding-number determination, as long as the cross section is available experimentally or theoretically. 

\begin{figure}
\begin{center}
\includegraphics[width=0.60\textwidth]{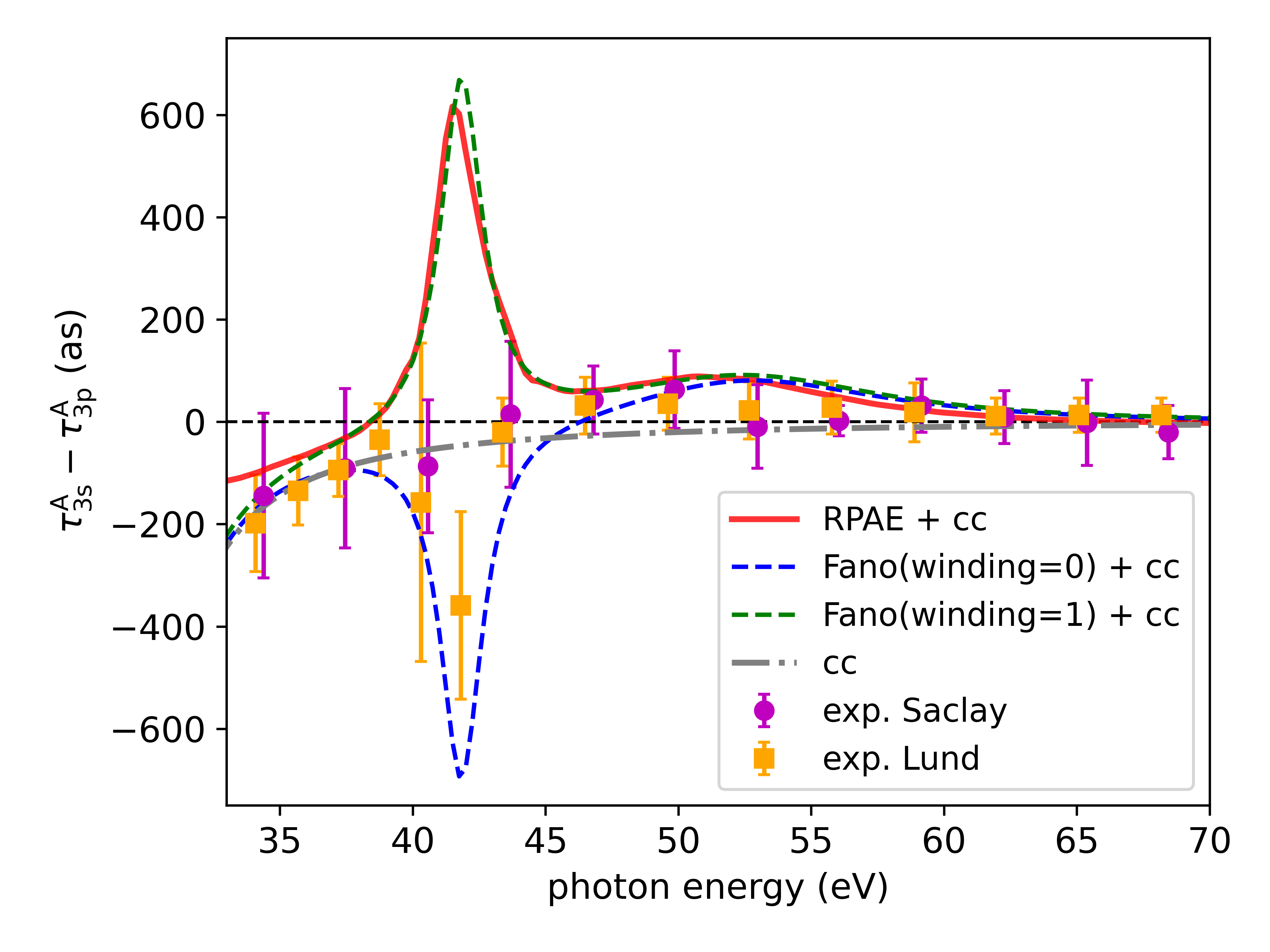}
\end{center}
\caption{Experimental \cite{alexandridi2021attosecond} and computed (RPAE, same as in figure \ref{fig:Cooper}) relative time delays between argon $3s$ and $3p$, compared with the time delay retrieved from the Fano parameters using equations (\ref{eq:tau_analytical}, \ref{eq:tau_analytical_winding_1}). The contribution from the continuum-continuum (cc) transition due to the dressing IR field in the RABBIT measurement is addressed by adding an additional cc time delay given in \cite{dahlstrom2014study}. \label{fig:Ar_Cooper_minimum} } 
\end{figure}

The validity of the relation between cross section and time delay for the CM of Ar $3p$ can be examined by the RABBIT experiment measuring the group delay of photorecombination upon high harmonic generation (HHG) \cite{schoun14a}, which can be physically understood as the inverse process of the photoionisation. As discussed in \cite{PhysRevA.91.063415}, the transition amplitudes of the two processes are equal and therefore feature the same time delay. The concurrence of the photoionisation cross section and photorecombination probability has been demonstrated by HHG studies \cite{woerner09a,shiner11a}. Figure \ref{fig:Cooper}(h) manifests the agreement of time delay between the experimental results and the RPAE calculation \cite{kheifets2023wigner} and the analytical model with Fano parameters. The experiment therefore verifies that the winding number of the transition amplitude for Ar $3p$ CM is 0, and as long as the winding number is determined, the time delay can be derived from the cross section variation at the CM.
On the other hand, theoretical calculations yield winding number of 1 for the Ar $3s$ CM, as the trajectory of the transition amplitude encircles the origin, as shown in figure \ref{fig:Cooper}(f). However, the RABBIT measurement reported in \cite{alexandridi2021attosecond} suggests that the time delay of Ar $3s$ CM is more negative than that of Ar $3p$, which contradicts the theoretical calculations. As compared in figure \ref{fig:Ar_Cooper_minimum}, with the fitted Fano parameters from the computed cross sections, assuming that Ar $3s$ has winding number of 1 at CM, the retrieved relative time delay between Ar $3s$ and $3p$ matches the computation, while assuming that Ar $3s$ has winding number of 0 at CM leads to a better agreement with the measured values which strikingly deviate from the theoretical prediction with the opposite sign. This indicates that the other pathways, e.g. the shake-up channels, shift the trajectory near the origin, so that the winding number changes from 1 to 0 and leads to a negative delay. 
We note that for the Xe $4d$ or the Ar $3p$ CM, the kinematic node appears only in one of the two photoionisation channels, which is offset by a finite contribution from the another channel, while for the Ar $3s$ CM, the $ 3s \rightarrow Ep $ channel is the only channel, if the target-state channel coupling is neglected. In this case, the trajectory exactly crosses the origin on the complex plane and leads to the discontinuity of the phase and thus a $\delta$-function time delay with arbitrary sign. Therefore, the finite time delay only comes from those couplings, and its sign depends on the detailed coupling terms. In fact, 
the subtle shift of the trajectory with $g = 0.046$ causes the topological change and thus flips the time delay and leads to a deviation of more than 1000 as.

\section{Conclusions}

In conclusion, we have introduced a Kramers-Kronig-like relation
between photoionisation cross sections and attosecond time delays,
which relies on the general properties of coherence and causality of
the transition amplitude. We have derived a unified time-delay formula
of Lorentzian and Fano lineshapes with constant and coherent
background, where both Breit-Wigner and Fano resonances correspond to
a pair of Lorentzian time-delay peaks with opposite signs. 
{We analysed several cases of the CM in valence shells of noble gas atoms. The relation is manifested by the excellent agreement of the retrieved time delays near CM of Xe $4d$ and Ar $3p$ from their cross sections, in comparison to the calculated time delay by RPAE.}
The LHT further
requires the topological property that the trajectory has winding
number of 0, which is particularly relevant for antiresonances where
the minimal cross section approaches zero. 
The theoretical calculations suggest that Xe $4d$ and Ar $3p$ have winding number of 0 at CM, whereas Ar $3s$ has winding number of 1 at CM.
The photorecombination experiment at Ar $3p$ CM \cite{schoun14a} shows good agreement with the computation and the analytical Fano model with winding number of 0. 
The latest experimental results on photoionisation \cite{alexandridi2021attosecond} rather hint at a negative time delay near the CM of Ar $3s$, which is at variance with the RPAE \cite{kheifets2013time} and TDLDA \cite{PhysRevA.91.063415} predictions and suggests a winding number of 0. So this particular case remains controversial and suggests a need for additional investigations. Our approach shows that this seemingly large change in delays can be cause by a small shift of the transition amplitude in the complex plane.

We note that our method is not restricted to atomic one-photon
processes, since similar resonances have been found in molecular
systems
\cite{huppert16a,heck2021attosecond,nandi20a,holzmeier2021influence}
and in two-photon transitions
\cite{jimenez2014modulation,jimenez2016two,argenti2017control,baykusheva17a}. Our result bridges two kinds of experiments in atomic and molecular
physics and broadens the understanding of attosecond processes.

\appendix

\section{Formulation of the Lorentzian peak and the Fano peak}
\label{sec:appendix_lorentz_fano}
The Lorentzian line shape is given by 
\begin{equation}
    \sigma = \sigma_{\rm a} \frac{1}{\epsilon^2+1} + \sigma_{\rm b} 
    = \sigma_{\rm b} \frac{\epsilon^2+1+{\sigma_{\rm a}/\sigma_{\rm b}}}{\epsilon^2+1} ~ .
\end{equation}
Compared with (\ref{eq:sigma_general}), we have $\mathfrak{S}=\sigma_{\rm b}$, $Q=0$, and $\gamma^2=1 + \sigma_{\rm a}/\sigma_{\rm b}$. For the Fano line shape, 
\begin{equation}
    \sigma = \sigma_{\rm a} \frac{{(\epsilon+q)}^2}{\epsilon^2+1} + \sigma_{\rm b}
    = (\sigma_{\rm a} + \sigma_{\rm b}) \frac{\epsilon^2+\frac{2q\epsilon+q^2+r}{r+1}}{\epsilon^2+1} ~ .
\end{equation}
Compared with (\ref{eq:sigma_general}), we have $\mathfrak{S}=\sigma_{\rm a} + \sigma_{\rm b}$, $2Q=\frac{2q}{r+1}$ and $Q^2+\gamma^2=\frac{q^2+r}{r+1}$, which leads to the expressions in the main text.
From (\ref{eq:sigma_general}) we have
\begin{eqnarray}
    \fl \frac{1}{\sigma}\frac{\partial \sigma}{\partial \epsilon}
    = \frac{2(\epsilon+Q)(\epsilon^2+1)-\left({(\epsilon+Q)}^2+\gamma^2\right)\cdot2\epsilon}{\left({(\epsilon+Q)}^2+\gamma^2\right)(\epsilon^2+1)}
    = \frac{2\left(\epsilon+Q\right)}{{\left(\epsilon+Q\right)}^2 + \gamma^2} - \frac{2\epsilon}{\epsilon^2+1}
\end{eqnarray}
which yields expression (\ref{eq:dsigma_dE_sigma}).

\section{Phase-jump calculation}
\label{sec:appendix_phase_jump}
For $\mathfrak{D}(\epsilon)$ in equation (\ref{eq:D_epsilon}), we have
\begin{eqnarray}
    {\rm Re}\{ \mathfrak{D}(\epsilon) \}/\sqrt{\mathfrak{S}} &&= \frac{\epsilon^2+Q\epsilon+\gamma}{\epsilon^2+1} \\
    {\rm Im}\{ \mathfrak{D}(\epsilon) \}/\sqrt{\mathfrak{S}} &&= \frac{(\gamma-1)\epsilon-Q}{\epsilon^2+1} ~ .
\end{eqnarray}
It can be verified that
\begin{eqnarray}
    {\left( \frac{{\rm Re}\{ \mathfrak{D}(\epsilon) \}}{\sqrt{\mathfrak{S}}} - \frac{\gamma+1}{2} \right)}^2 + {\left( \frac{{\rm Im}\{ \mathfrak{D}(\epsilon) \}}{\sqrt{\mathfrak{S}}} + \frac{Q}{2} \right)}^2
    = \frac{1}{4}\left( {(\gamma-1)}^2+Q^2 \right) .
\end{eqnarray}
Hence, the trajectory of $\mathfrak{D}(\epsilon)$ is a circle with the radius of
$\frac{\sqrt{\mathfrak{S}}}{2}\sqrt{{(\gamma - 1)}^2 + Q^2}$ centred
at $\left( \sqrt{\mathfrak{S}}\frac{\gamma+1}{2},
-\sqrt{\mathfrak{S}}\frac{Q}{2} \right)$. Since the origin is $\frac{\sqrt{\mathfrak{S}}}{2}\sqrt{{(\gamma + 1)}^2 + Q^2}$ away from the centre of the circle, the tangent segment from the origin is $\sqrt{\mathfrak{S}}\sqrt{\gamma}$, and the angle between the two tangents yields 
\begin{eqnarray}
    {|\Delta \phi|}_{\max} &&= \max(\phi(\epsilon)) - \min(\phi(\epsilon)) 
    = 2 {\cos}^{-1} \left( \frac{2\sqrt{\gamma}}{\sqrt{{(\gamma+1)}^{2} + Q^2}} \right).
\end{eqnarray}
At $\epsilon \rightarrow -\infty$, the starting point of the trajectory approaches $(\sqrt{\mathfrak{S}}, 0)$, which is left to the centre when $\gamma > 1$ while right to the centre when $\gamma < 1$. Since the trajectory evolves counter-clockwise, the phase between the two tangents increases when $\gamma > 0$ and decreases when $\gamma < 0$. Therefore, the phase jump for $\gamma \neq 1$ can be expressed as
\begin{equation}
    {\Delta \phi}_{\max} = 2 {\rm sgn}(\gamma-1) {\cos}^{-1} \left( \frac{2\sqrt{\gamma}}{\sqrt{{(\gamma+1)}^{2} + Q^2}} \right).
\end{equation}
For $\gamma = 1$, it corresponds to a symmetric phase variation, where the maximal (minimal) phase is at $\epsilon=0$ while the minimal (maximal) phase is at $\epsilon=\pm\infty$, and 
\begin{eqnarray}
    \phi^{\gamma=1}(0) - \phi^{\gamma=1}(\infty) &&= -2 {\rm sgn}(Q) {\cos}^{-1} \left( \frac{1}{\sqrt{1 + {(Q/2)}^2}} \right) \nonumber \\
    &&= -2 {\tan}^{-1}(Q/2) ~ .
\end{eqnarray}
In the special case of $Q=0$, namely the Lorentzian peak, the expression can be simplified to
\begin{equation}
    {\Delta \phi}^{Q=0}_{\max} = 4 {\tan}^{-1}(\sqrt{\gamma}) - \pi
\end{equation}
which is positive when $\gamma > 1$ (maximum in cross section) and is negative when $\gamma < 1$ (minimum in cross section). It approaches $+{\rm \pi}$ and $-{\rm \pi}$ for $\gamma \rightarrow +\infty$ ($\sigma_{\rm b}/\sigma_{\rm a} \rightarrow -1$) and $\gamma \rightarrow 0$ ($\sigma_{\rm b}/\sigma_{\rm a} \rightarrow 0$), respectively.

\section{RPAE calculation}
\label{sec:appendix_RPAE}
The RPAE calculations were performed using the {\sc atom} program suite
\cite{AC97}. For argon, the correlations between the three optically
allowed transitions $3s\to Ep$ and  $3p\to Es/Ed$  were taken
into account. For Xe $4d$, the five transitions $5s\to Ep$,   $4p\to
Es/Ed$ and $4d\to Ep/Ef$ were included. The cross section in the
polarization direction was was evaluated from the total photoionization
cross section and the angular anisotropy parameter as
$\sigma(1+\beta)$. The length gauge results were used for both
$\sigma$ and $\beta$. The time delay in the polarization direction was
evaluated from the photoionization amplitude as in
\cite{kheifets2013time}. Summation over the magnetic projections in
the ground state was reduced to $m_i=0$.  

\section{Fano parameters for cross sections near Cooper minima}
\label{sec:appendix_Fano_CM}
The cross section of a Fano resonance can be conventionally expressed as \cite{fano1961effects,fano1965line,kossmann1988new}:
\begin{equation}
    \sigma_{\rm F}(\epsilon) = \sigma_0(\epsilon)
    \left( \rho^2 \frac{{(q+\epsilon)}^2}{1+\epsilon^2}+1-\rho^2 \right)
\end{equation}
where $\epsilon = (E-E_{\rm r})/(\Gamma/2)$, and $\rho^2$ is known as the correlation coefficient, and it is linked to equation (\ref{eq:fano_peak}) by letting $\sigma_{\rm a} = \rho^2 \sigma_0$, $\sigma_{\rm b} = (1-\rho^2) \sigma_0$, $r = \sigma_{\rm b}/\sigma_{\rm a} = 1/{\rho^2} - 1$, and approximating $r$ as a constant across the CM. From the RPAE calculation, the Fano parameters are 
$E_{\rm r} = 105.46~{\rm eV}$, $\Gamma = 64.0282~{\rm eV}$, $\sigma_0 = 6.26653~{\rm Mb}$, $q = -2.45287$, $\rho^2 = 0.990963$ for Xe $4d$ CM, 
$E_{\rm r} = 29.2918~{\rm eV}$, $\Gamma = 18.468~{\rm eV}$, $\sigma_0 = 8.62031~{\rm Mb}$, $q = -2.67896$, $\rho^2 = 0.934333$ for Ar $3p$ CM, and
$E_{\rm r} = 31.4118~{\rm eV}$, $\Gamma = 19.6322~{\rm eV}$, $\sigma_0 = 0.567772~{\rm Mb}$, $q = -1.06661$, $\rho^2 = 0.995518$ for Ar $3s$ CM, 
respectively, which correspond to the fitted curves in figure \ref{fig:Cooper}.

\section*{Acknowledgements}
J.-B.Ji acknowledges the funding from the ETH grant 41-20-2. M.H.'s work was funded by the European Union’s Horizon 2020 research and innovation programme under Marie Skłodowska-Curie agreement grant No. 801459, FP-RESOMUS. J.-B.Ji thanks S. Luo (Jilin University), J. O. Richardson (ETH Z{\"u}rich) and R. R. Lucchese (Lawrence Berkeley National Laboratory) for discussions.

\section*{Author contributions statement}
J.-B.J. derived the formulae with support of A.S.K., M.H., and K.U. A.S.K. performed the RPAE calculation and the fitting. J.-B.J., K.U., and H.J.W. conceived the study. M.H., K.U., and H.J.W. supervised its realization. All authors discussed the results and wrote the paper.

\section*{Competing interests statement}
All co-authors have seen and agree with the contents of the manuscript and there is no financial interest to report.

\section*{References}
\bibliographystyle{iopart-num}
\bibliography{attobib,KK_relations,references}

\end{document}